\documentclass[fleqn,usenatbib]{mnras}

\usepackage{newtxtext,newtxmath}
\usepackage{color}

\usepackage[T1]{fontenc}
\usepackage{ae,aecompl}

\usepackage{graphicx}	
\usepackage{amsmath}	
\usepackage{amssymb}	

\usepackage{multirow}

\newcommand{\Msolar}{\mbox{\,$\rm M_{\odot}$}}        
\newcommand{\Rsolar}{\mbox{\,$\rm R_{\odot}$}}        
\newcommand{\Lsolar}{\mbox{\,$\rm L_{\odot}$}}        

  \newcommand{\Teff}{\mbox{\,\em T$_{\rm eff}$}}         

  \newcommand{\yr}{\,\mbox{yr}}                          
  \def\simge{\mathrel{\raise1.16pt\hbox{$>$}\kern-7.0pt
    \lower3.06pt\hbox{{$\scriptstyle \sim$}}}}           
  \def\simle{\mathrel{\raise1.16pt\hbox{$<$}\kern-7.0pt
    \lower3.06pt\hbox{{$\scriptstyle \sim$}}}}           
\title[Pulsation in post-common-envelope binary stars]{Post-common-envelope binary stars, radiative levitation, and blue large-amplitude pulsators}

\author[C.~M.~Byrne \& C.~S.~Jeffery]{
Conor M. Byrne$^{1,2}$\thanks{E-mail: conor.byrne@armagh.ac.uk (CMB)}
and C. Simon Jeffery$^{1,2}$
\\
$^{1}$Armagh Observatory and Planetarium, College Hill, Armagh BT61 9DG, UK\\
$^{2}$School of Physics, Trinity College Dublin, College Green, Dublin 2, Ireland\\
}

\date{Last updated XXXX; in original form YYYY}

\pubyear{2018}

\begin{document}
\label{firstpage}
\pagerange{\pageref{firstpage}--\pageref{lastpage}}
\maketitle

\begin{abstract}
Following the discovery of blue large-amplitude pulsators (BLAPs), single star evolution models of post red giant branch stars that have undergone a common envelope (CE) ejection in the form of a high mass loss rate have been constructed and analysed for pulsation stability. The effects of atomic diffusion, particularly radiative levitation, have been examined. Two principal models were considered, being post-CE stars of 0.31 and 0.46 \Msolar. Such stars are likely, in turn, to become either low-mass helium white dwarfs or core helium-burning extreme horizontal-branch stars. The inclusion of radiative levitation leads to opacity driven pulsations in both types of post-CE object when their effective temperatures are comparable to those of BLAPs, with similar periods. The extent of the instability region for models in these simulations, which are not in thermal balance, is larger than that found for static models, in agreement with previous theory. By comparing to observations, and making some simple evolutionary assumptions, we conclude the 0.31 \Msolar\ star is the more likely candidate for BLAPs. The rate of period change is negative for both cases, so the origin of BLAPs with positive rates of period change remain uncertain.
\end{abstract}




\section{Introduction}
Pulsational instability is a behaviour observed in stars all across the Hertzsprung-Russell diagram, from  classical pulsators such as the Cepheid and Mira variables, to  more recent discoveries such as the blue large-amplitude pulsators (BLAPs), variable stars found in the OGLE survey \cite{Pietrukowicz17}. These objects show brightness variations of 0.2--0.4 magnitudes with periods of 20--40 minutes. Model atmosphere fits to spectroscopic observations indicate the stars have effective temperatures $T_{\rm{eff}}\approx 30\,000\,\rm{K}$ and surface gravities $\log(g)\approx 4.6$. \cite{Pietrukowicz17} report a surface helium mass fraction of 0.52 for one of the BLAP prototype object, OGLE-BLAP-001. The high surface gravity and large helium abundance suggest an unusual evolutionary history. The fact that these hydrogen deficient objects pulsate is perhaps unsurprising given that opacity-driven radial pulsations are  excited  in many regions of the luminosity--\Teff\ plane when the the damping influence of hydrogen is diminished \citep{JefferySaio16}. 

The surface gravity and effective temperature of OGLE-BLAP-001 place it below the main sequence, suggesting that a significant amount of mass may have been lost during its evolution \cite{Pietrukowicz17}. This is one of the similarities with extreme horizontal branch (EHB) stars. The latter are core helium burning stars with low mass hydrogen envelopes. They have $T_{\rm eff} \approx 20\,000 - 40\,000\,$K, $5.4 \leq \log(g) \leq 6.0$ and many have helium rich surfaces. They include the classical subdwarf B stars, which have helium-poor surfaces, as well as various hydrogen-rich and and helium-rich  subdwarf O  (sdO and He-sdO) stars. Proposed formation mechanisms for EHB include common envelope ejection of a red giant with a core of sufficient mass to begin fusion of helium. A similar mechanism acting on a red giant with an inert 0.31\Msolar\ core could produce an object with properties comparable to those observed for BLAPs, according to envelope models of the pulsations computed by \cite{Pietrukowicz17}. 

Most hot subdwarfs have a peculiar surface composition. Although most have helium-depleted surfaces, surface helium abundances can vary from almost 0 to 100 percent. A few show overabundances of elements such as lead and zirconium \citep{Naslim11}. Some hot subdwarfs show helium-rich surfaces comparable to that of OGLE-BLAP-001 with a helium mass fraction of 0.52. Like BLAPs, many hot subdwarfs are known to show brightness variations due to pulsations. Hot subdwarfs that oscillate in p-modes have periods of 2--9 minutes and surface temperatures of $28\,000 \leq T_{\rm{eff}}\slash K \leq 35\,000$. The pulsations in hot subdwarfs are driven by an opacity bump ($\kappa$-mechanism) due to the presence of iron and nickel, which are enhanced in the outer layers through the process of radiative levitation. These pulsations were predicted theoretically by \cite{Charpinet96,Fontaine03} and discovered observationally by \cite{Kilkenny97,Green03}.


When in hydrostatic and radiative equilibrium, atomic species may migrate upwards or downwards within a star by diffusion wherever a suitable gradient exists. Examples include  concentration diffusion and thermal diffusion.
Radiative levitation is where radiation pressure in the stellar interior imparts different forces on different ions according to on their electron structure. High ionization states of iron-group elements have a dense line spectrum and thus absorb a large fraction of  incident radiation, producing a nett upward force at temperatures around $2\times10^5$K. 
The downward diffusion of heavier elements referred to as gravitational settling is due to a microscopic imbalance in the gravitational and electric forces acting on ions. 
The competition between these processes causes ions to float (or levitate) at positions where the differential forces acting on different ions are in equilibrium. 

Recent work has investigated the effects of atomic diffusion and radiative levitation in particular on the evolution of a post-common envelope star as it evolves from the
red giant branch (RGB) to the extreme horizontal branch (EHB) following the onset of helium burning in the core \citep{Byrne18}. Here we expand the analysis of these simulations to include searches for pulsations in pre-EHB models. We also investigate post-common envelope models of lower mass, to investigate the extent of any instability strip which may exist, as well as the driving mechanism behind it. The analysis  focuses on specific models in the evolutionary sequence which have surface gravities and temperatures comparable to those of BLAPs in order to test the possibility that the direct progenitors of either hot subdwarfs or low mass white dwarfs could correspond to BLAPs.

\begin{figure*}
\centering
\includegraphics[width=0.9\textwidth]{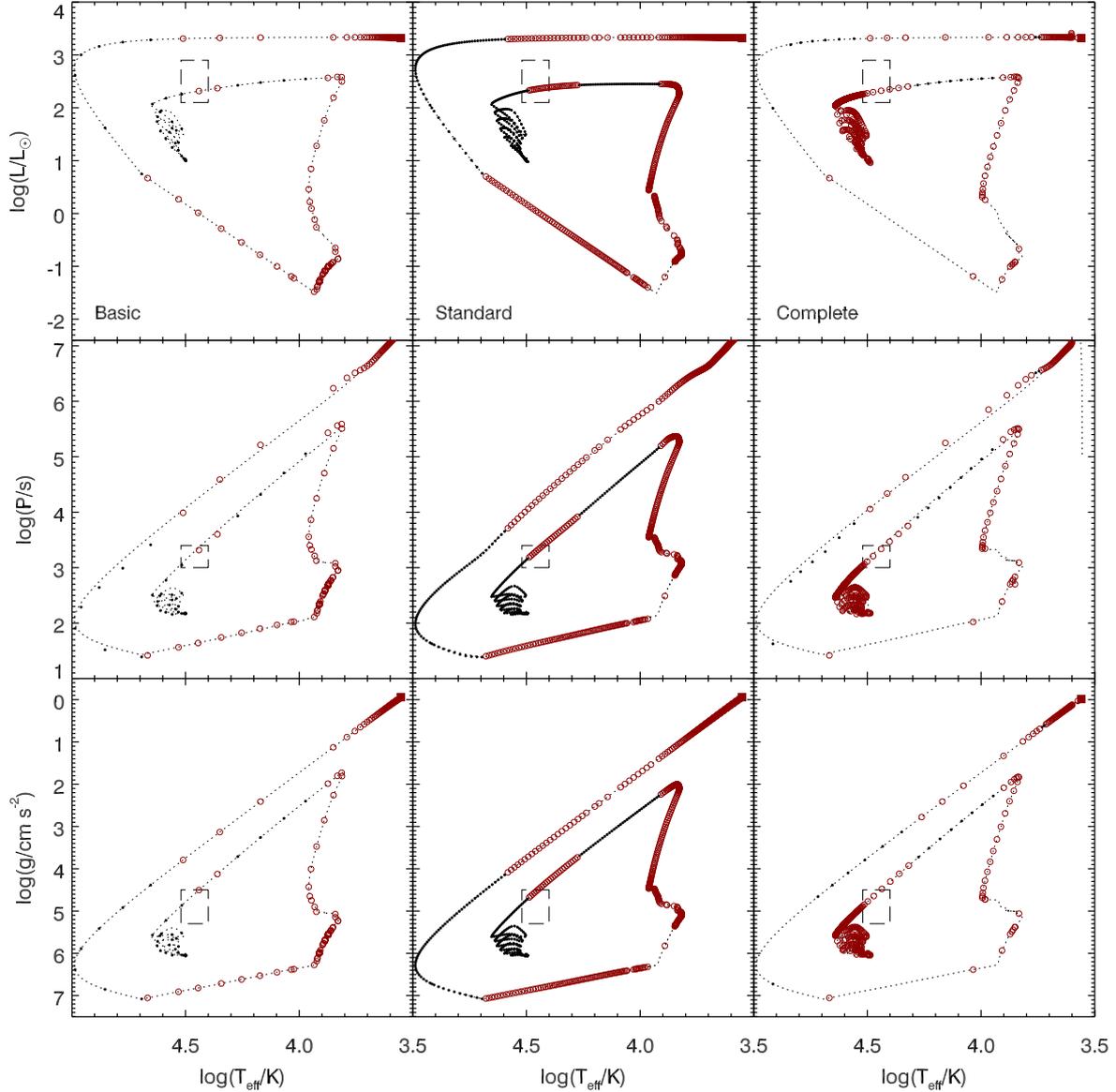}
\caption{Evolution of luminosity (top row), fundamental period (middle row) and surface gravity (bottom row) as a function of effective temperature for the basic, standard and complete models of the 0.46\Msolar\ post-common-envelope models. Solid black dots indicate models found to be stable, while the open red circles indicate those with an unstable fundamental mode. The dashed rectangles indicate the approximate range of values of these parameters within which BLAPs have been found as reported by \protect\cite{Pietrukowicz17}. Interpolation of a smoothly varying pulsation constant over the evolution means that not all points in the period-effective temperature diagram fall precisely on the evolution track fitted to the data.}
\label{fig:evolutionsdb}
\end{figure*}

\section{Methods}
Models of stellar evolution were computed using the computer program {\sc{mesa}} \citep[revision 7624]{Paxton11,Paxton13,Paxton15,Paxton18}. 
The methods adopted and parameter space explored closely follow the previous paper by \cite{Byrne18}.

The evolutionary tracks of some of the pre-EHB models calculated by \cite{Byrne18} pass close to the region of the luminosity -- effective temperature diagram where BLAPs have been identified. 
In this region, pre-EHB models match several of the observed properties of BLAPs, including inflated envelopes and helium-enriched surfaces. In this paper, we consider in detail one model from \cite{Byrne18}, namely model 3, having an initial (zero-age main sequence) mass of 1 \Msolar, and a pre-EHB mass of 0.46 \Msolar\ after simulated common-envelope ejection close to the tip of the RGB.

As described in \cite{Byrne18}, key model parameters include the mixing length $\alpha_{\rm{MLT}}=1.9$, following \cite{Stancliffe16}, metallicity  $Z = 0.02$ with the mixture of \cite{Grevesse98}, the Schwarzchild criterion for convection and a helium mass fraction $Y =0.28$. More particularly, this model shows  strong hydrogen-shell burning and flash-driven mixing which succeed in removing nearly all the surface hydrogen. The product is a helium main-sequence star, which might be identified spectroscopically as a helium-rich subdwarf O star (He-sdO).  

Another possible structure for a BLAP, proposed by \cite{Pietrukowicz17} and partially explored by \cite{Romero18}, is a star with a small helium core of about 0.31\Msolar\ that has been stripped of its envelope. 
For this work, a 0.31\Msolar\ post-common envelope star was produced using the same method as for the hot subdwarf stars in \cite{Byrne18}. That is, by implementing a large mass loss rate ($\dot{M}=10^{-3}\,\Msolar\,\rm{yr}^{-1}$) to strip most of the envelope, leaving only a small hydrogen envelope of $\sim 3\times10^{-3}\,\Msolar$.
This was allowed to evolve until it becomes a white dwarf with $\log L/\Lsolar\,< -2$. The zero-age main sequence star had a mass of 1 \Msolar, $Y=0.28$ and $Z=0.02$.

Common envelope evolution remains a poorly understood phase of evolution. However, given that it is believed to happen on a dynamical timescale, the approach taken in this work is quite reasonable, as any detailed analysis of the structure is only done when the time since envelope ejection is longer than the thermal timescale of the star.

\begin{figure*}
\centering
\includegraphics[width=0.9\textwidth]{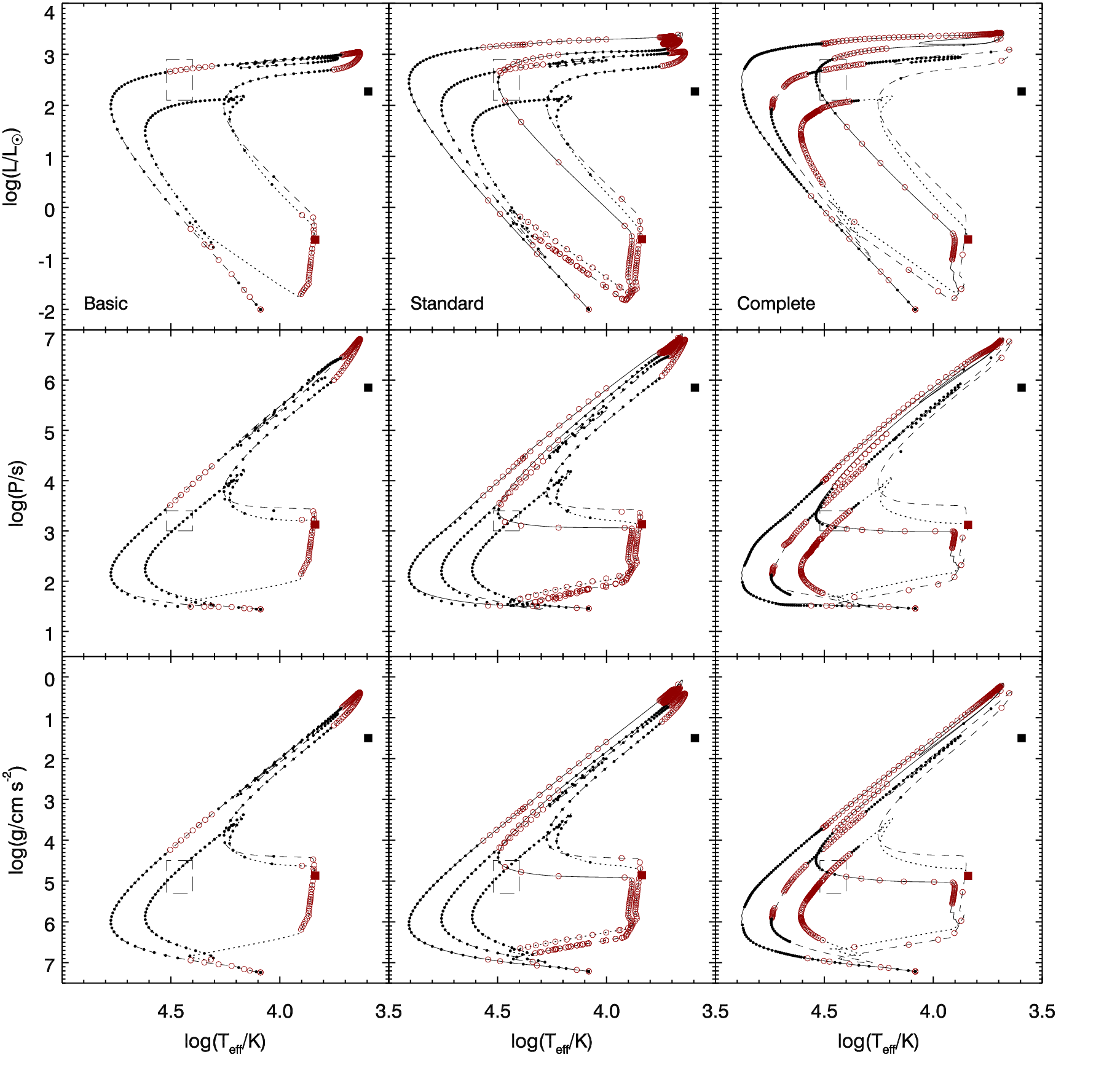}
\caption{As Fig.~\protect\ref{fig:evolutionsdb}, but for the 0.31\Msolar\ post-common-envelope models. The black squares indicate the location of the red giant model before envelope ejection, while the corresponding red squares indicate the first snapshot following envelope ejection. To assist with clarity, the first `loop' in the diagram is indicated by a dotted line, while the second and third loops are represented by dashed and solid lines respectively.}
\label{fig:evolutionblap}
\end{figure*}

\subsection{Diffusion and radiative levitation}
The primary objective is to examine the effect of atomic diffusion, and radiative levitation in particular, on the chemical structure of the stellar envelope and the pulsation stability of the models as they pass through different phases of evolution.   
{\sc{mesa}} uses the approach of \cite{Thoul94} to solve Burgers diffusion equations \citep{Burgers69}. 
Modifications to this approach which are required in order to compute radiative levitation are described by \cite{Hu11}. 
This allows calculation of the radiative forces on all ions of specified elements for the chemical mixture in each of the outer layers of the star. 
In this study, radiative accelerations are calculated for all layers of the star with a temperature less than $10^7\,\rm{K}$.
In order to isolate the effects of radiative levitation and following \cite{Byrne18}, the evolution was calculated without diffusion (hereafter referred to as basic models), with thermal diffusion, concentration diffusion and gravitational settling (standard models), and with thermal diffusion, concentration diffusion, gravitational settling and radiative levitation (complete models).

\subsection{Pulsation}

To analyse the stability of the models, the oscillation code {\sc{gyre}} was used \citep{Townsend13}. 
Outputs from {\sc{mesa}} can be configured as input to {\sc{gyre}} which allows for straightforward analysis of pulsation stability.

An adiabatic analysis was carried out to determine the eigenfrequencies of the model, followed by a non-adiabatic analysis to investigate the stability of the modes identified. As  BLAPs are known to be large-amplitude pulsators, only radial modes ($l=0$) were investigated, although the excitation of other low-degree modes ($l\leq2$) is possible.  A frequency scan was chosen to identify the fundamental mode and the first few radial overtones ($k\simle5$). The sign of the imaginary component of the eigenfrequency $\omega$, determined from the non-adiabatic analysis was used to determine the stability of the mode. The sign convention  in {\sc{gyre}} is that eigenfunctions are of the form
\begin{equation}
y(t) = A\exp[-i\omega t].
\label{eq:eigen}
\end{equation}
Thus, if the imaginary component of the eigenfrequency is positive (negative), this mode is unstable (stable) and there is driving (damping) of that mode. In this paper, models described as stable/unstable are defined solely from the fundamental mode ($l=0, k=0$). 

\section{Results}
We examine first the evolution of an EHB star progenitor, followed by that of a low-mass WD progenitor. 
The primary difference between the two cases is that, while both suffer a series of hydrogen shell flashes which lead to one or more loops in the HR (or $g-\Teff$) diagram, only the first ignites helium in the core. Models for post-CEE pre-WD evolution have been presented elsewhere \citep[e.g.]{Romero18}, but none have included the effects of radiative levitation. 

\subsection[0.46 Msun post-common-envelope pre-extreme horizontal branch star]{0.46\Msolar\ post-common-envelope pre-extreme horizontal branch star}

Our EHB progenitor model corresponds to model 3 of \cite{Byrne18}. This is a 1\Msolar\ red giant star near the tip of the red giant branch, which has had almost all of the envelope removed to leave a 0.46\Msolar\ star.  This model was chosen as it undergoes a hydrogen shell flash, leading to a helium rich surface for at least part of its evolution, thus mimicking the surface properties of OGLE-BLAP-001.
Fig.~\ref{fig:evolutionsdb} shows the evolution of luminosity, fundamental period and surface gravity against effective temperature for the 0.46\Msolar\ post-CEE star, for each of the diffusion options. Each symbol along the evolution tracks represents a {\sc{mesa}} snapshot of the model which was analysed for pulsations. The small black dots represent models where the fundamental  mode is stable. The red open circles indicate models that have an unstable fundamental mode and would be expected to pulsate. The dashed rectangle indicates the region of parameter space in which BLAPs are typically found, as defined by \cite{Pietrukowicz17}.

The results demonstrate three regions of instability during the transition from common envelope ejection to the horizontal branch. The first in the constant luminosity phase immediately succeeding the common envelope ejection. This region appears reasonably similar regardless of the choice of diffusion physics. The second begins in the low luminosity phase preceding the first helium flash (and corresponding hydrogen shell flash) and persists until the model reaches maximum post-flash luminosity. This phase is also seen in all three models. The final phase of instability begins when the model returns to an effective temperature of around $20\,000\,\rm{K}$ ($\log(\Teff) = 4.3$). In the basic and standard models, stability returns once the temperature exceeds around $32\,000\,\rm{K}$ ($\log(\Teff) = 4.5$), while instability persists in the complete model all the way to the extreme horizontal branch. The time taken to cross the instability strip between $20\,000\,\rm{K}$ and $32\,000\,\rm{K}$ (which roughly corresponds to the region of interest in which a BLAP may be observed) is approximately $1.5\times10^4\,\rm{yr}$.

\begin{figure*}
\centering
\includegraphics[width=0.9\textwidth]{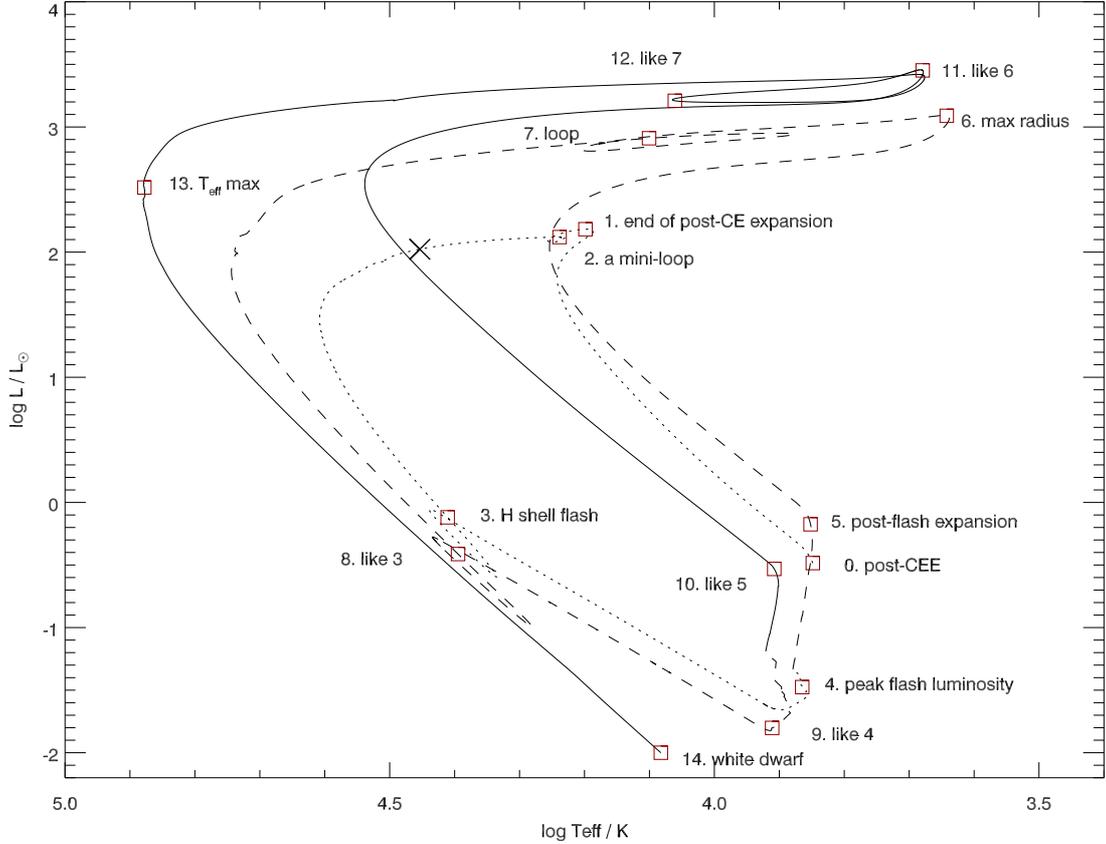}
\caption{Evolution track of a 0.31\Msolar\ post-common-envelope model including radiative levitation. The square symbols indicate the location of the numbered points. For clarity, the continuous track is represented by three line-styles, from dotted, through dashed, to continuous. The cross indicates the model with parameters comparable to a BLAP which was chosen for further investigation.}
\label{f:track31}
\end{figure*}

\subsection[0.31 Msun post-common-envelope pre-extreme horizontal branch star]{0.31\Msolar\ post-common-envelope pre-white dwarf}

Fig.~\ref{fig:evolutionblap} shows the same results as Fig.~\ref{fig:evolutionsdb}, but for the model with a core mass of 0.31\Msolar. 
One significant difference between these models is the number of hydrogen shell flashes which occur. In the basic model, only one hydrogen shell flash is seen, while both the standard and complete models undergo two hydrogen shell flashes, completing two loops in the diagram. Each loop occurs at a higher temperature and luminosity than the previous loop. The reason for the larger number of flashes is likely due to diffusion creating a hydrogen-abundance gradient such that shell flashes are less efficient, and leaving hydrogen-rich layers sufficiently massive to trigger additional flashes. As with the EHB progenitor, pulsation modes are unstable at a number of points. Similarly these occur during the constant luminosity phases of the evolution and in the aftermath of the hydrogen shell flash. However, in the case of the complete model, a large portion of the first (innermost) loop is found to be unstable, and it is this phase of evolution when the surface gravity, effective temperature and fundamental pulsation period tend to best match that of the BLAPs.

Since the track in Fig.~\ref{fig:evolutionblap} is complicated and novel, it is expanded and annotated in Fig.~\ref{f:track31}. Key points are identified as follows.  
The point labelled 0 indicates the initial state of the model after the rapid removal of mass from the envelope at a rate of $10^{-3}\,$\Msolar$\,$yr$^{-1}$. The model then expands in order to return to equilibrium, reaching maximum radius at point 1, after about $5\times10^2\,$yr. In doing so, it overshoots its equilibrium radius leading to another phase of contraction and expansions, which manifests itself as the loop at point 2, taking $6\times10^4\,$yr. The star then evolves towards the white dwarf cooling track, taking $10^7\,$yr to reach point 3. At this point, the hydrogen luminosity begins to increase rapidly, indicating the onset of the hydrogen shell flash, which reaches peak luminosity when the model is at point 4 in the figure. After this, the strength of the flash starts to decline; energy produced during the flash is transported into the envelope, which heats and expands. The transition from point 3 to point 6 takes approximately 100 yr. At point 6, maximum expansion is achieved and the star begins to contract towards white dwarf cooling track again. A small loop occurs, labelled as point 7. This appears to be related to a slight increase in the hydrogen luminosity, and possibly a redistribution of energy within the stellar envelope. The behaviour of the model from points 8 to 12 is similar to that from points 3 to 7, with comparable timescales. The star eventually cools to become a white dwarf at 14, taking $3.1\times10^4\,$yr to contract from point 11 to maximum temperature at point 13, and a further $2.24\times10^8\,$yr to cool to point 14, where the star has a luminosity of $10^{-2}$\Lsolar.

Table~\ref{t:track31} shows the elapsed time and time between various points on the evolution track shown in Fig.~\ref{f:track31}. 

\begin{table*}
\centering
\caption{Features in Fig~\ref{f:track31} with a brief description of the events related to the numbered points, the time elapsed since the previous numbered point, the duration of the `loops' (where relevant), total time elapsed since the end of the CE ejection phase and a brief description of the behaviour at these points.}
\label{t:track31}
\begin{tabular}{rllll}
Feature & $\rm{T}_{n}-\rm{T}_{n-1}$/yrs & $\rm{T}_{loop}$/yrs & $\rm{T}_{tot}$/yrs & Notes                               \\ \hline
0  & 0                   & 0                & 0                & Initial model properties after CE mass loss completed       \\
1  & $4.16\times10^2$    & -                & $4.16\times10^2$ & End of post-CE expansion phase                              \\
2  & $4.75\times10^2$    & $6.12\times10^4$ & $6.23\times10^4$ & A small loop as the model settles into an equilibrium state \\
3  & $1.03\times10^7$    & -                & $1.04\times10^7$ & Onset of the first hydrogen shell flash                     \\
4  & $8.81\times10^0$    & -                & $1.04\times10^7$ & Peak in hydrogen flash luminosity                           \\
5  & $2.86\times10^1$    & -                & $1.04\times10^7$ & Star expands in response to flash                           \\
6  & $5.71\times10^1$    & -                & $1.04\times10^7$ & End of post-flash expansion                                 \\
7  & $5.30\times10^1$    & $3.91\times10^2$ & $1.04\times10^7$ & A post-flash loop                                           \\
8  & $2.93\times10^7$    & -                & $3.97\times10^7$ & Onset of the second hydrogen shell flash                    \\
9  & $1.10\times10^{-1}$ & -                & $3.97\times10^7$ & Peak in the flash luminosity, as 4                        \\
10 & $3.79\times10^0$    & -                & $3.97\times10^7$ & As 5                                                        \\
11 & $3.17\times10^1$    & -                & $3.97\times10^7$ & As 6                                                        \\
12 & -                   & $1.20\times10^2$ & $3.97\times10^7$ & As 7                                                        \\
13 & $3.08\times10^3$    & -                & $3.97\times10^7$ & Temperature maximum \\
14 & $2.24\times10^8$    & -                & $2.64\times10^8$ & White dwarf with $\log L / \Lsolar = -2$                         \\
\end{tabular}
\end{table*}

A clear result is that the inclusion of atomic diffusion and radiative levitation has a major impact, not only on the chemical structure and hence, potentially, on the pulsation properties, but also on the overall evolution of the models from envelope ejection through to the white dwarf phase.
To compare with the 0.46\Msolar\ model, the time taken for this model to cross the temperature range of $20\,000\,\rm{K}$ to $32\,000\,\rm{K}$ is around $8\times10^{5}\,\rm{yrs}$. Thus, purely from an evolutionary timescale standpoint, a BLAP is more likely to be a 0.31\Msolar\ post-common envelope star than a 0.46\Msolar\ star, since the lower mass object spends about 50 times as long crossing the region of interest.

\subsection{Driving}

\begin{figure}
\centering
\includegraphics[width=0.4\textwidth]{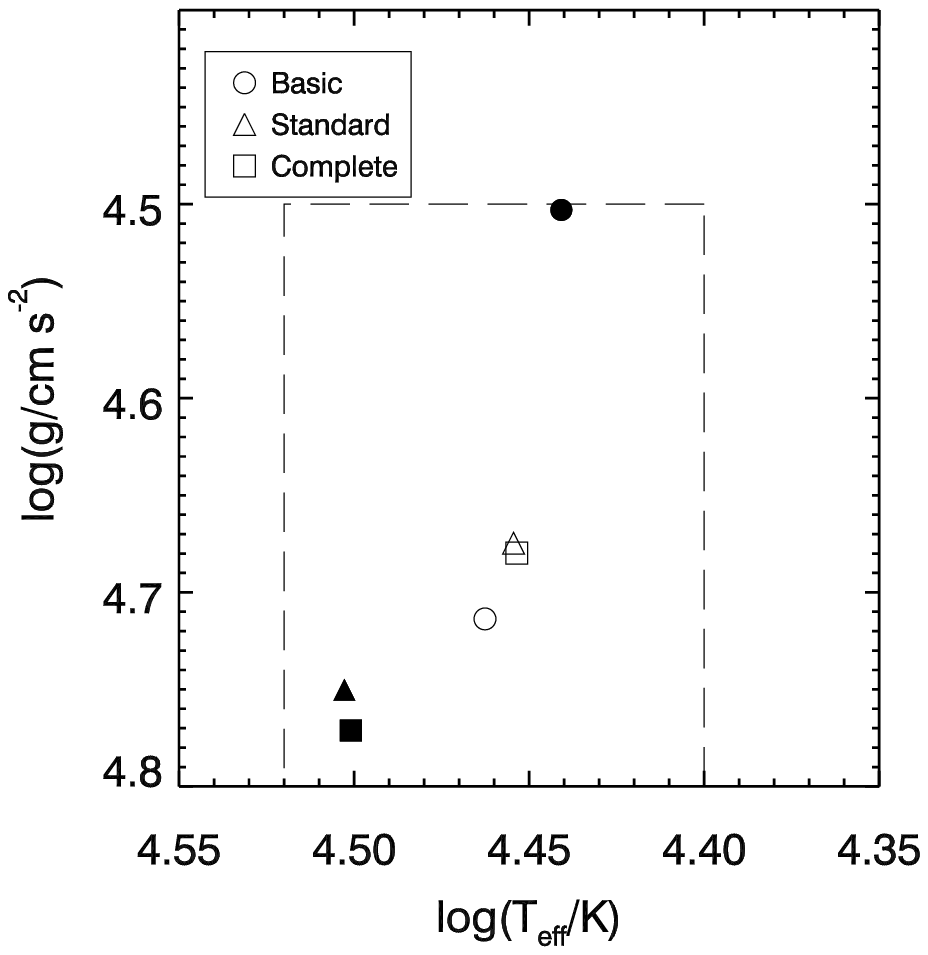}
\caption{Models chosen from the different evolution tracks as candidate BLAP objects on a surface gravity-effective temperature diagram. The filled symbols represent models of the 0.46\Msolar\ post-CEE star while the open symbols represent models of the 0.31\Msolar\ post-CEE star. The dashed box indicates part of the boundary region in which BLAPs have been identified, and extends to $\log(g)\approx5.3$.}
\label{fig:closeup}
\end{figure}

\begin{figure}
\centering
\includegraphics[width=0.38\textwidth]{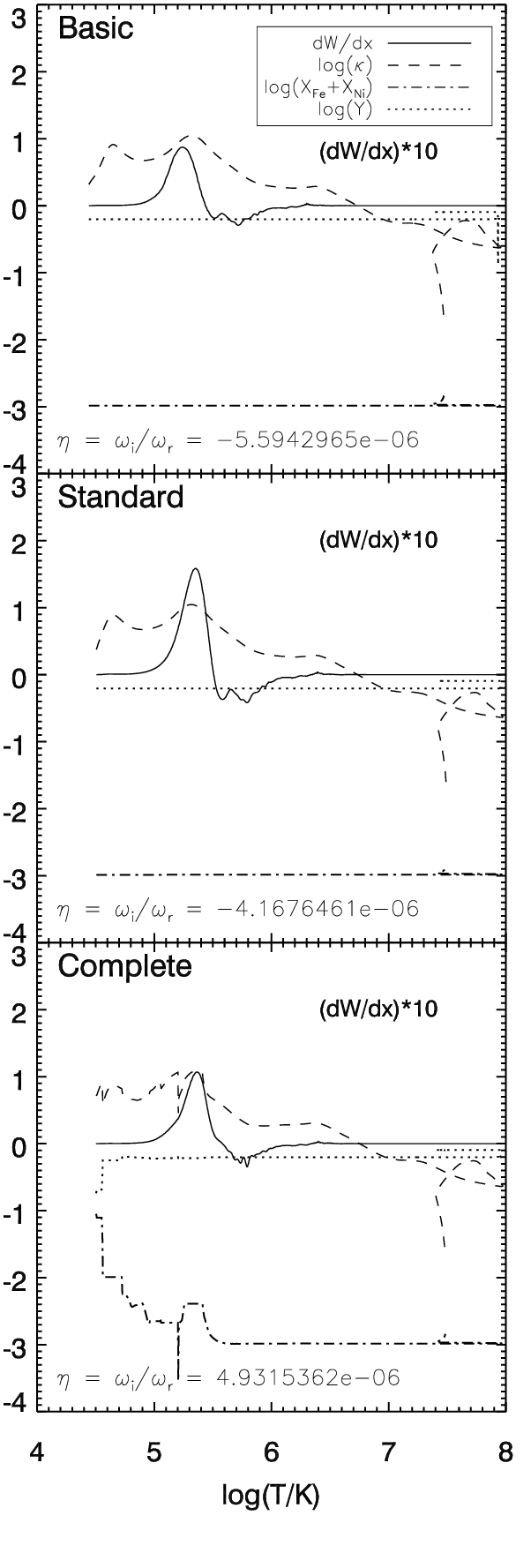}
\caption{Opacity ($\log \kappa / {\rm cm^{-1}}$) and derivative of the work function of the fundamental mode (${\rm d}W/{\rm d}x$, in units of G$\,\rm{M}_*^2\,\rm{R}_*$) as a function of temperature for the 0.46\Msolar\ pre-EHB models from Table~\ref{tab:blapmodels}. Where indicated, ${\rm d}W/{\rm d}x$ has been multiplied by 10 for visibility. The mass fraction of helium and the combined mass fraction of iron and nickel are also shown. All functions are multiply valued for $\log T /{\rm K} > 7.5$ owing to the temperature inversion caused by neutrino cooling in the degenerate core.}
\label{fig:drivesdb}
\end{figure}

\begin{figure}
\centering
\includegraphics[width=0.40\textwidth]{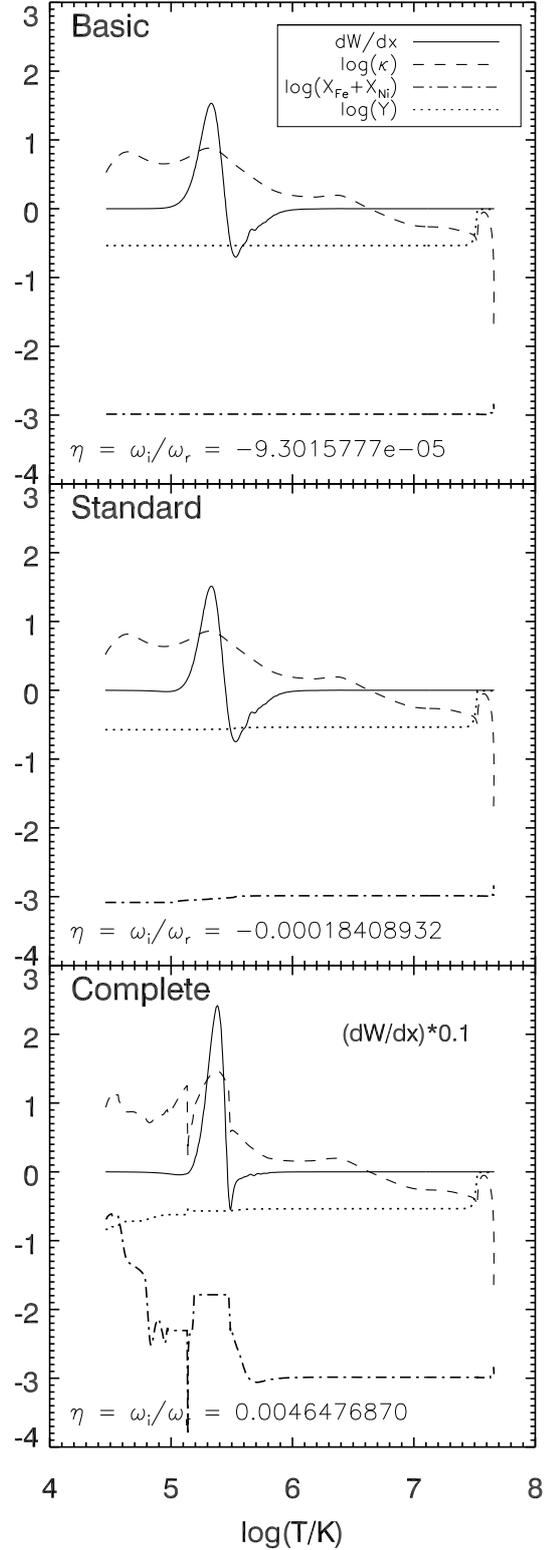}
\caption{Same as Fig.~\protect\ref{fig:drivesdb} for the 0.31 \Msolar\ pre-WD models from Table~\ref{tab:blapmodels}. 
Where indicated, ${\rm d}W/{\rm d}x$ has been multiplied by 0.1 for visibility alongside the other parameters in the diagram.}
\label{fig:driveblap}
\end{figure}

In order to identify if any of these models could produce BLAP behaviour, individual models from each evolution track falling in the BLAP parameter range were selected (Fig.~\ref{fig:closeup}, Table~\ref{tab:blapmodels}). Figs.~\ref{fig:drivesdb} and \ref{fig:driveblap} show the logarithm of opacity ($\log(\kappa)$) and the derivative of the work function (${\rm d}W/{\rm d}x$) as a function of interior temperature for the pre-EHB and pre-WD models respectively. The mass fraction of helium and combined mass fraction of iron and nickel are also included.

\begin{table*}
\centering
\caption{Properties of models selected for detailed analysis, as shown in Fig.~\protect\ref{fig:closeup}.}
\label{tab:blapmodels}
\begin{tabular}{cccccccccc}
Model         & Physics  & $\frac{\rm{M}}{\Msolar}$      & $\frac{\rm{M}_{\rm{Core}}}{\Msolar}$& $\log\left(\frac{\rm{R}_*}{\Rsolar}\right)$    & $\log\left(\frac{\rm{L}_*}{\Lsolar}\right)$ & $\log\left(\frac{g}{\rm{cm}\,\rm{s}^{-2}}\right)$ & $\log\left(\frac{T_{\rm{eff}}}{\rm{K}}\right)$ & $\log\left(\frac{P_0}{\rm{s}}\right)$ & Stability \\ \hline
\multirow{3}{*}{pre-EHB} & Basic    & 0.46215 & 0.46215 & -0.2002 & 2.3163 & 4.5030 & 4.4408 & 3.0445 & Stable    \\
                         & Standard & 0.46215 & 0.46215 & -0.3237 & 2.3171 & 4.7502 & 4.5028 & 3.1309 & Stable    \\
                         & Complete & 0.46215 & 0.46215 & -0.3416 & 2.2738 & 4.7712 & 4.5009 & 3.1096 & Unstable  \\ \hline
\multirow{3}{*}{pre-WD}  & Basic    & 0.31045 & 0.30776 & -0.3919 & 2.0201 & 4.7137 & 4.4626 & 3.1169 & Stable    \\
                         & Standard & 0.31045 & 0.30772 & -0.3724 & 2.0264 & 4.6747 & 4.4544 & 3.1465 & Stable    \\
                         & Complete & 0.31045 & 0.30774 & -0.3727 & 2.0223 & 4.6798 & 4.4536 & 3.1236 & Unstable  \\ \hline
\end{tabular}
\end{table*}

These figures show that the driving zones in the models (the location of the peak in $\rm{d}W/\rm{d}x$) coincide with the location of the opacity maximum related to the ionisation of iron and nickel at $\log(T)\simeq5.3$. This is a clear indication that for models with unstable modes, the pulsations observed in these models are driven by the $\kappa$-mechanism arising from the opacity of iron-group elements. It is particularly apparent in the case of the 0.31\Msolar\ model, that the size of the opacity bump is significantly larger when radiative accelerations are computed. This is due to the accumulation of iron and nickel in this region of the envelope, which illustrates the important role played by radiative levitation. Additionally, the fundamental modes of the horizontal branch models are only unstable when radiative levitation is included, as shown in Fig.~\ref{fig:evolutionsdb}. This is indicative that radiative levitation plays a key role in causing sufficient amounts of iron to accumulate in the envelope of the star to drive pulsations. This result is not surprising, as it has been widely shown that the presence of enhanced levels of iron and nickel in the envelope of hot subdwarfs is needed to drive the pulsations \citep[][e.g]{Charpinet96,JefferySaio07,Fontaine08,Michaud11,Hu11}.
Furthermore, \cite{Romero18} suggested that iron and nickel opacity was the driving mechanism for BLAPs. They used an artificial uniform enhancement of metallicity to replicate the effects of radiative levitation in a simple manner. The more detailed radiative levitation calculations carried out in this work agree with the suggestion that iron and nickel opacity is responsible for the pulsations seen in BLAPs. 

\subsection{Rate of period change}

Observed period changes provide vital clues about other changes in a star, particularly one which may be evolving quickly. The observations of \cite{Pietrukowicz17} report rates of period change, $\dot{\Pi}$, in the range $-3 \times 10^{-7}\,\rm{yr}^{-1} \le$ $\dot{\Pi}$ $\le +8 \times 10^{-7}\,\rm{yr}^{-1}$, where $\dot{\Pi}$ is defined as
\begin{equation}
\dot{\Pi} = \frac{\Delta P}{\Delta t} \frac{1}{P}
\label{eq:pdot}
\end{equation}
and where $\Delta P$ is the change in period in a time interval $\Delta t$. 
\cite{Romero18} report values of $\dot{\Pi} \approx \pm10^{-7}-10^{-5}\,\rm{yr}^{-1}$  for their template model.

With the inclusion of the effects of radiative levitation, the rate of period change derived for our 0.31\Msolar\ model with radiative levitation in this work, identified as the `complete' 0.31\Msolar\ model in Fig,.~\ref{fig:closeup} and Table~\ref{tab:blapmodels}, we find a value of $\dot{\Pi}\simeq -3\times10^{-6}\,\yr^{-1}$, which is comparable in magnitude to those of the \cite{Romero18} models and the \cite{Pietrukowicz17} observations. 

A puzzle for both of the latter is that both negative and positive values of $\dot{\Pi}$ are reported. Generally speaking, a pulsation period 
will decrease (negative $\dot{\Pi}$) in a star which is contracting, and increase in a star which is expanding. 
It is not entirely clear from  \cite{Romero18} how values of $\dot{\Pi}$  were computed for their template model, but since all of their models are for contracting stars, only negative values should be expected. 
The challenge provided by the \cite{Pietrukowicz17} observations is that they imply BLAPs include both contracting and expanding stars.

Since our model is on the contracting stage of the evolution track (between points 2 and 3 on Fig. 3), we obtain a negative value of  $\dot{\Pi}$.  Another part of the track passes close to the BLAP zone (points 10 to 11 on Fig. 3) during an expansion phase, but with such a short lifetime (31 yr, Table 1) that (a) it would be an unlikely observation and (b) $\dot{\Pi}$ would be orders of magnitude larger than any observed. 

The most likely source of positive values of $\dot{\Pi}$ is an expanding star. From the models studied for this paper, the closest match to this is provided by loops associated with off-centre helium ignition flashes in the 0.46 \Msolar\ star, prior to core helium ignition, as seen in Fig.~\ref{fig:evolutionsdb}. While the temperature and luminosity of the models presented here do not agree with the observations by \cite{Pietrukowicz17}, the exact locations of these loops are dependent on the models used and the assumptions around the behaviour of common envelope ejection, such as the amount of hydrogen envelope remaining after common envelope ejection among other things. Thus it is possible that stars in this phase of evolution could be the source of BLAPs with a positive value of $\dot{\Pi}$.

\section{Discussion}

Fig.~\ref{f:nmodes} compares the  instability of models shown in Figs.~\ref{fig:evolutionsdb} and \ref{fig:evolutionblap} with the survey of instability in hydrogen-deficient stellar envelopes  by \cite{JefferySaio16}. 
The high-luminosity tracks, show blue edges in approximately the same locations. 
Given that the computational tools are completely independent, this validates the methods. 
The most significant differences are instability in the models approaching the EHB and in hottest pre-WD models each time they approach a maximum in effective temperature.

There are several significant differences between the models of \cite{JefferySaio16} and those presented here. The former consider homogeneous stellar envelopes in full radiative and hydrostatic equilibrium. Such models work well for early-type stars which evolve slowly. The mean-molecular weight in the envelope, which governs the equation of state, is dominated by the hydrogen and helium abundances. The opacity in the major driving zones is governed by these and the metal abundances. The energy flux (luminosity) is assumed to be constant throughout the envelope. By ensuring that the composition reflects the mean-molecular weight through the envelope, and the distribution of metals in the potential driving zones, pulsation stability can be explored in large parameter spaces.  

The envelope models shown in Fig.~\ref{f:nmodes} differ from the current {\sc{mesa}} models in two  crucial ways. 
The first is that, as a result of radiative levitation, metal abundances in the driving zones are very substantially enhanced relative to the initial metallicity ($Z$), which
is similar to that adopted in the envelope models. The radiative enhancement 
can be simulated in the envelope models  by increasing the iron and nickel contribution to the opacity by, say, a factor 10 \citep[cf.]{JefferySaio06b}. 
The bottom row of  Fig.~\ref{f:nmodes} shows how such an increase also increases the parameter space (for hot stars) in which the models are unstable. As shown in Figs.~\ref{fig:drivesdb} and~\ref{fig:driveblap}, the combined mass fraction of iron and nickel, $\log(X_{\rm{Fe}}+X_{\rm{Ni}})$,  in the driving region is around --2.4 for the chosen evolutionary phase of the 0.46 \Msolar\ model which closely resembles a BLAP, while $\log(X_{\rm{Fe}}+X_{\rm{Ni}})=-1.8$ for the 0.31 \Msolar\ model.

The second difference is that the {\sc{mesa}} evolution models  correctly include the gravothermal term in the energy equation (${\rm d}l/{\rm d}m = \epsilon - T{\rm d}S/{\rm d}t$) which is non-negligible during phases of rapid evolution such as those illustrated in Figs.~\ref{fig:evolutionsdb} and \ref{fig:evolutionblap}. 
It is not  clear how much this  affects the constant flux assumption (${\rm d}l/{\rm d}m = 0$) used for the \cite{JefferySaio16} stability survey, but it is an interesting 
question of wider significance to objects sometimes referred to as 'bloated' stars.  

\subsection*{Bloated stars}

Terms such as 'bloated' or 'inflated' are sometimes used to refer to envelopes of stars which are not in full hydrostatic, radiative and thermal equilibrium. 
This may be a consequence of either a sudden increase in the core luminosity (core or shell flash), or a loss of radiative support when a nuclear energy source is depleted. 
The implication is that flux entering the envelope from below is substantially different from that leaving the surface. 
The temperature and density structure must therefore adjust until the two quantities are equal; the envelope must either contract to  make good a deficit, or expand to absorb an excess. 
In the case of contraction, the process will normally be quasi-static and changes take place on the thermal timescale of the envelope. 
Expansion phases may also be in thermal equilibrium, but can sometimes be driven dynamically by more explosive events (e.g.  shallow shell flashes). 
The latter can give rise to two types of 'bloated' envelope; first, the expanding envelope driven dynamically by a nuclear ignition and, second, a contracting envelope after the nuclear energy source has switched off. The fact that the model is either expanding or contracting on short timescales indicates that it is not in equilibrium. 
In either case, the structure may be quite different from that of an  envelope in which ${\rm d}l/{\rm d}m = 0$.
The question is how this affects pulsation stability,  which occurs on a dynamical timescale.

The question can be addressed for an individual case by looking at the magnitude of the gravothermal term in models pertaining to BLAPs. Fig.~\ref{fig:dldm} shows the magnitude of the gravothermal term in the sense  ${\rm d}l/{\rm d}m$ through the envelopes of two models.
Note that the sign of ${\rm d}l/{\rm d}m$ changes in the region of the iron-nickel opacity bump. 
So although the bulk of the envelope is contracting, and hence converting gravitational potential energy to heat (${\rm d}l/{\rm d}m > 0$), heat is being re-absorbed within the opacity bump. 
The reason is that the opacity bump can be attributed primarily to the ionization of K- and L-shell electrons in the iron and nickel atoms \cite{OP1,OP2}. 
The same ionization produces an increase in the local specific heat capacity. 
As the envelope contracts, all layers of the star contract and heat; as the ionization zone moves outward into a new layer, it locally traps heat (${\rm d}l/{\rm d}m < 0$).
Once ionization is complete, the same layer, which is now beneath the ionization zone, continues to contract and release heat as before (${\rm d}l/{\rm d}m > 0$).

The non-zero value of the gravothermal term has several consequences for the structure and stability of the envelope. The most obvious is that it directly affects the radiative gradient through the transport equation (Equation~\ref{eq:transport})
\begin{equation}
\frac{{\rm d}T}{{\rm d}m}=\frac{3\kappa \rho^2 l }{4 a c T^3}
\label{eq:transport}
\end{equation}
via the flux $l$. The gradient is further affected through the opacity $\kappa$ when radiative acceleration operates to concentrate material in locations where its own specific opacity is high.   
The perturbation to the radiative gradient will propagate through the equation of state to the pressure and density, and hence to the overall stability.  

\begin{figure*}
\begin{center}
\includegraphics[width=0.45\textwidth]{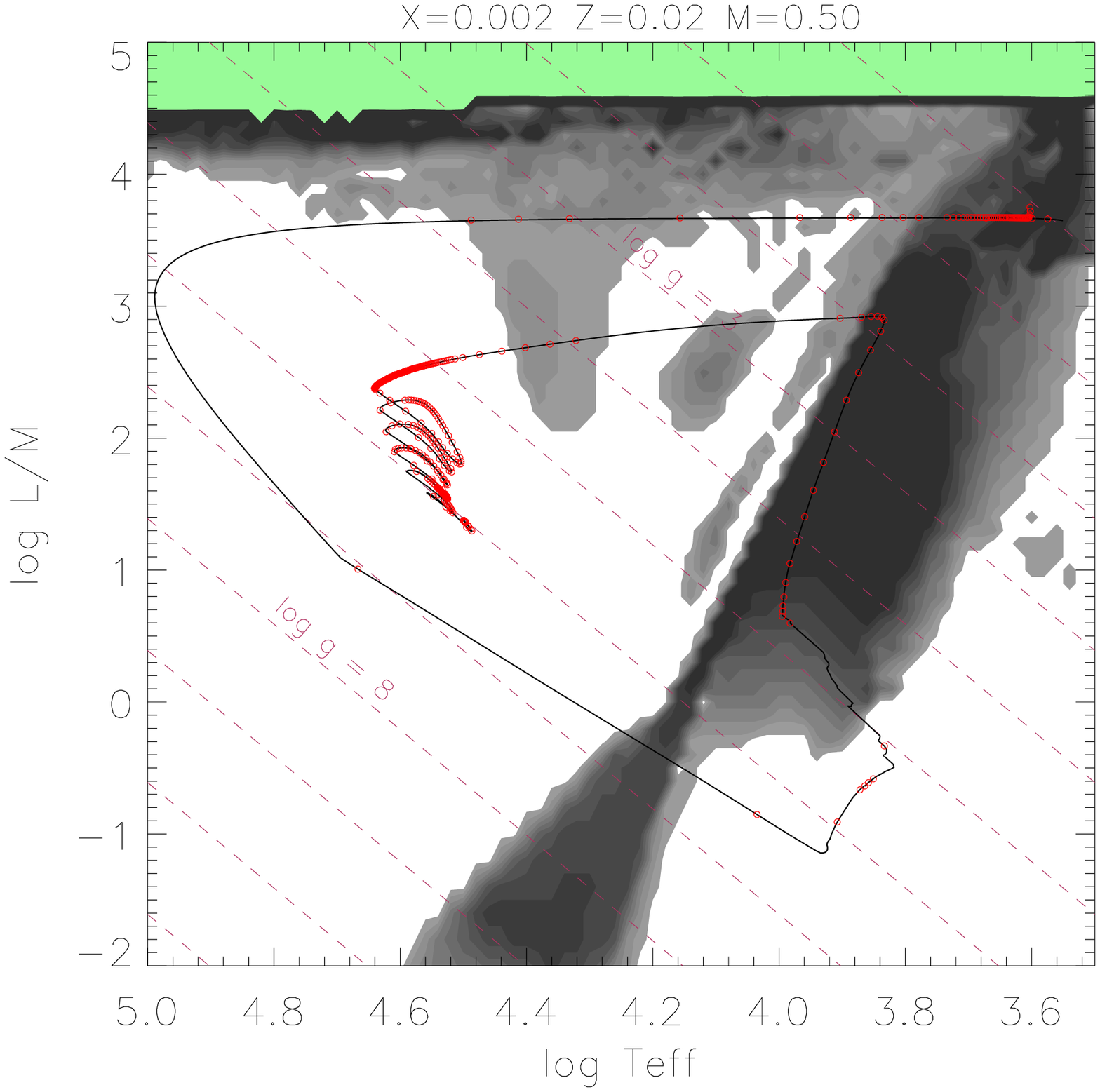}
\includegraphics[width=0.45\textwidth]{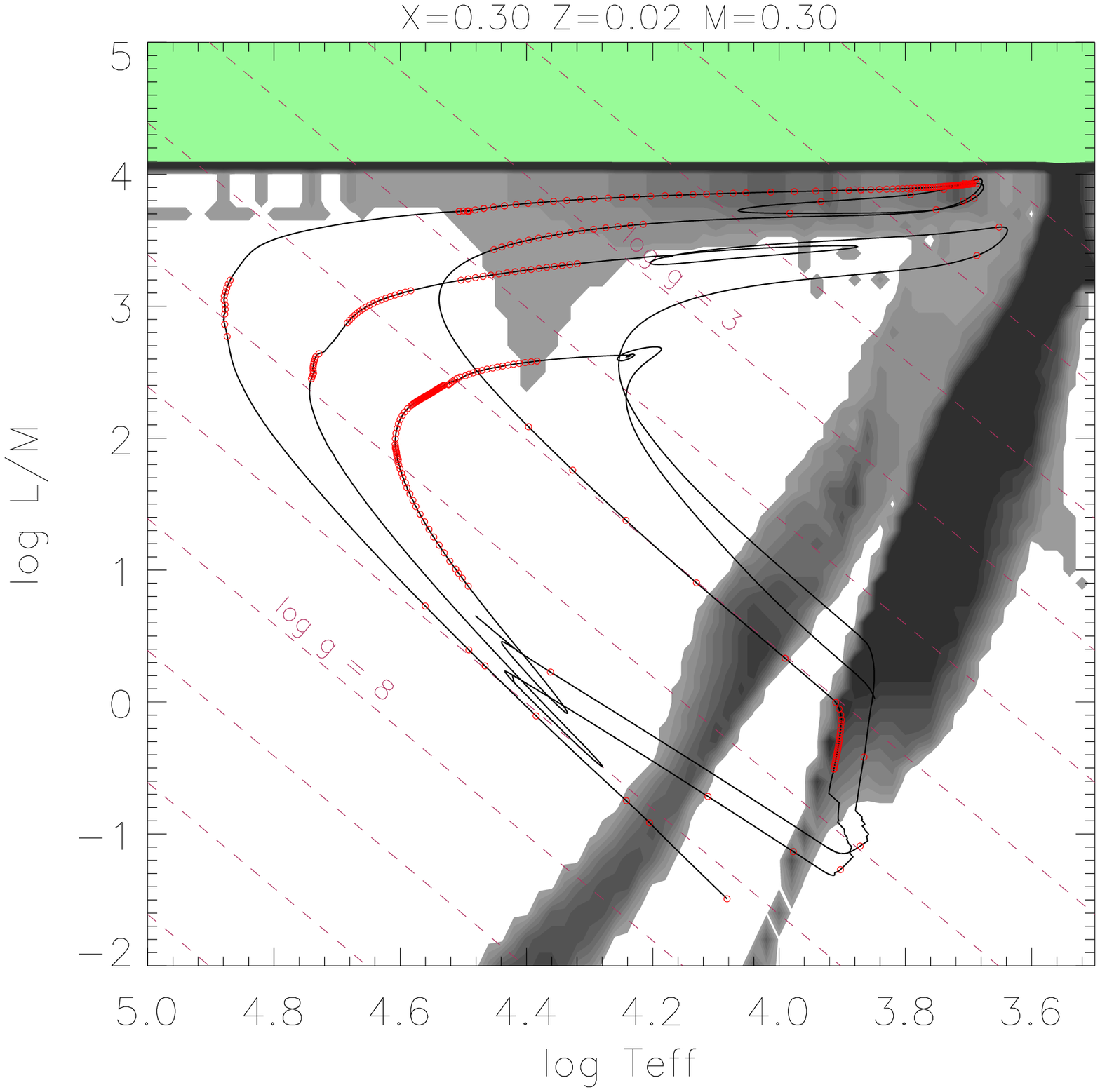}\\
\includegraphics[width=0.45\textwidth]{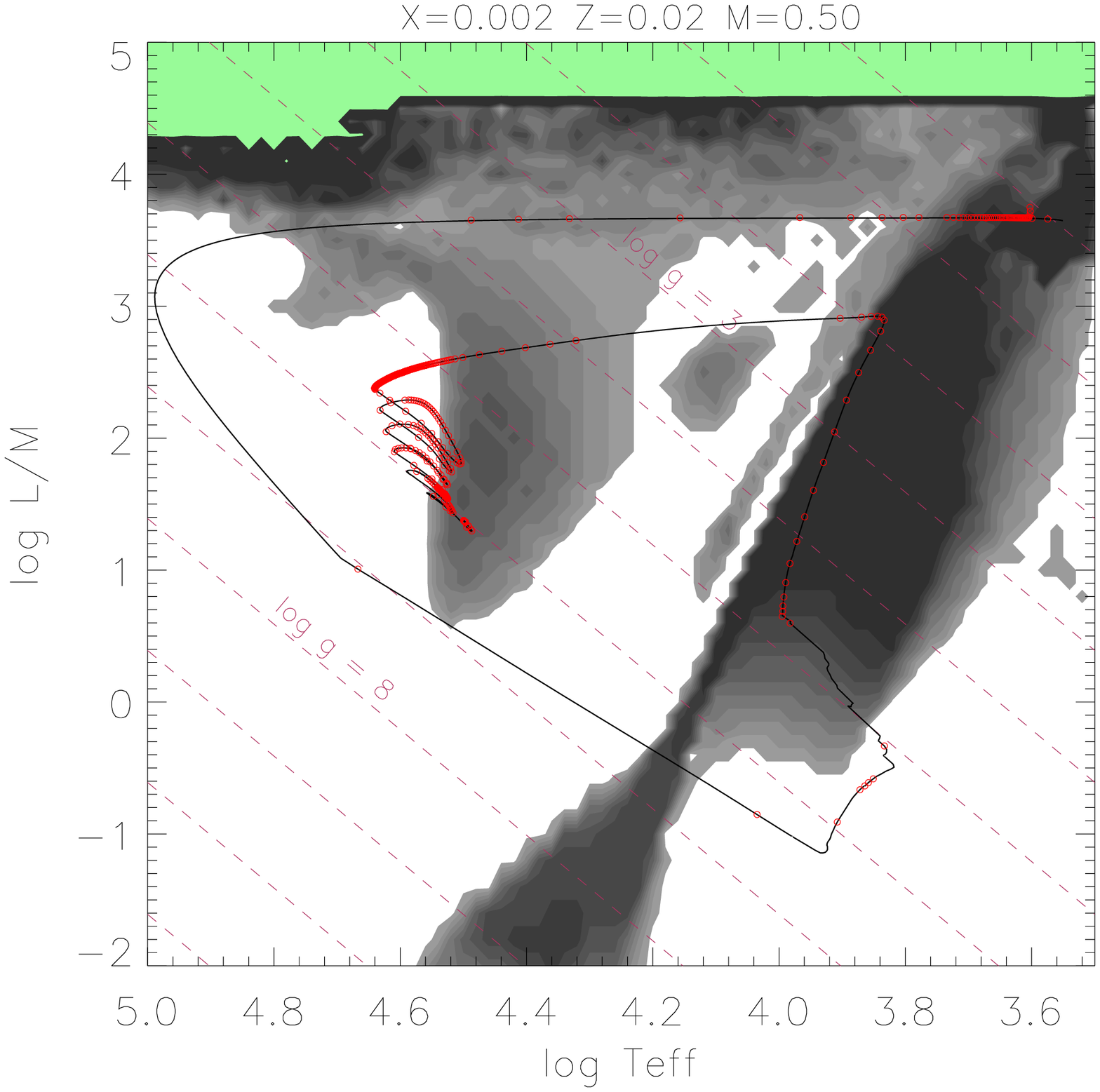}
\includegraphics[width=0.45\textwidth]{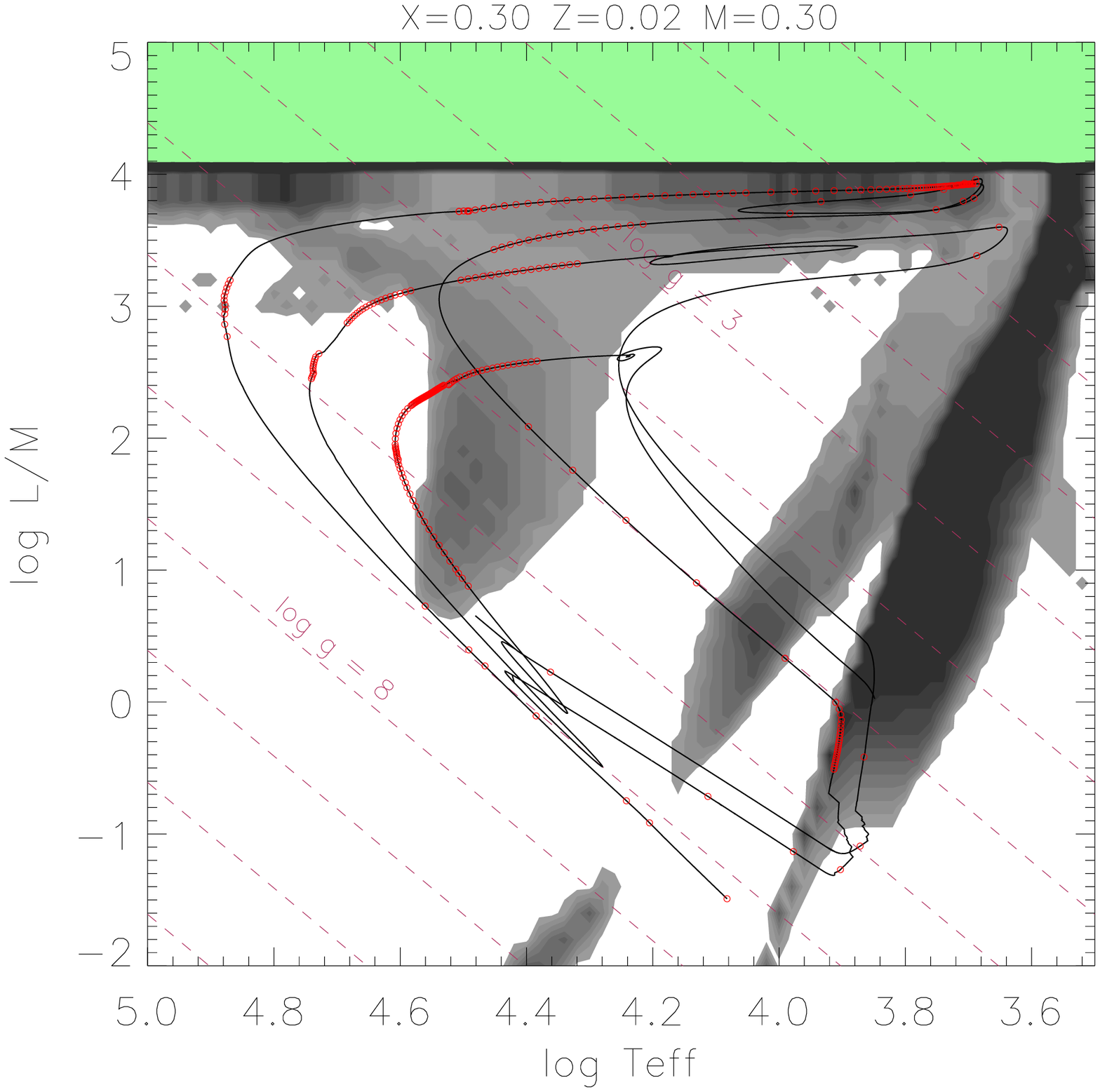}
\caption[Unstable modes and low-mass post-CEE evolution]
{Evolution tracks for post-CEE stars with radiative levitation superimposed on the loci of pulsationally unstable homogeneous envelopes obtained by \citet{JefferySaio16}, with   compositions and masses as labelled. \Teff\ is in degrees kelvin; the luminosity and mass are in solar units. The left panels show the track (solid line) of a 0.46\Msolar\ pre-EHB star (cf. Fig.\,1) and the right panels show the track of 0.31 \Msolar\ pre-WD  (cf. Fig.\,2). Red circles represent pulsationally unstable models on the evolution track of post-CEE models. 
The grey-scale contour plot represents the {\it number} of unstable radial modes, with the lightest shade marking the instability boundary
(one unstable mode), and the darkest shade representing ten or more unstable modes. 
Broken diagonal lines represent contours of constant surface gravity at  $\delta \log g  = 1$ ($g$ labelled in ${\rm cm s^{-2}}$). 
Pale green denotes regions where envelope models do not converge. 
The top row assumes a scaled-solar metallicity with $Z=0.02$. 
The bottom row assumes the same metal abundances but with iron and nickel increased by a factor of 10 \citep{JefferySaio06b}. 
}
\label{f:nmodes}
\end{center}
\end{figure*}

\begin{figure*}
\begin{center}
\includegraphics[width=0.45\textwidth]{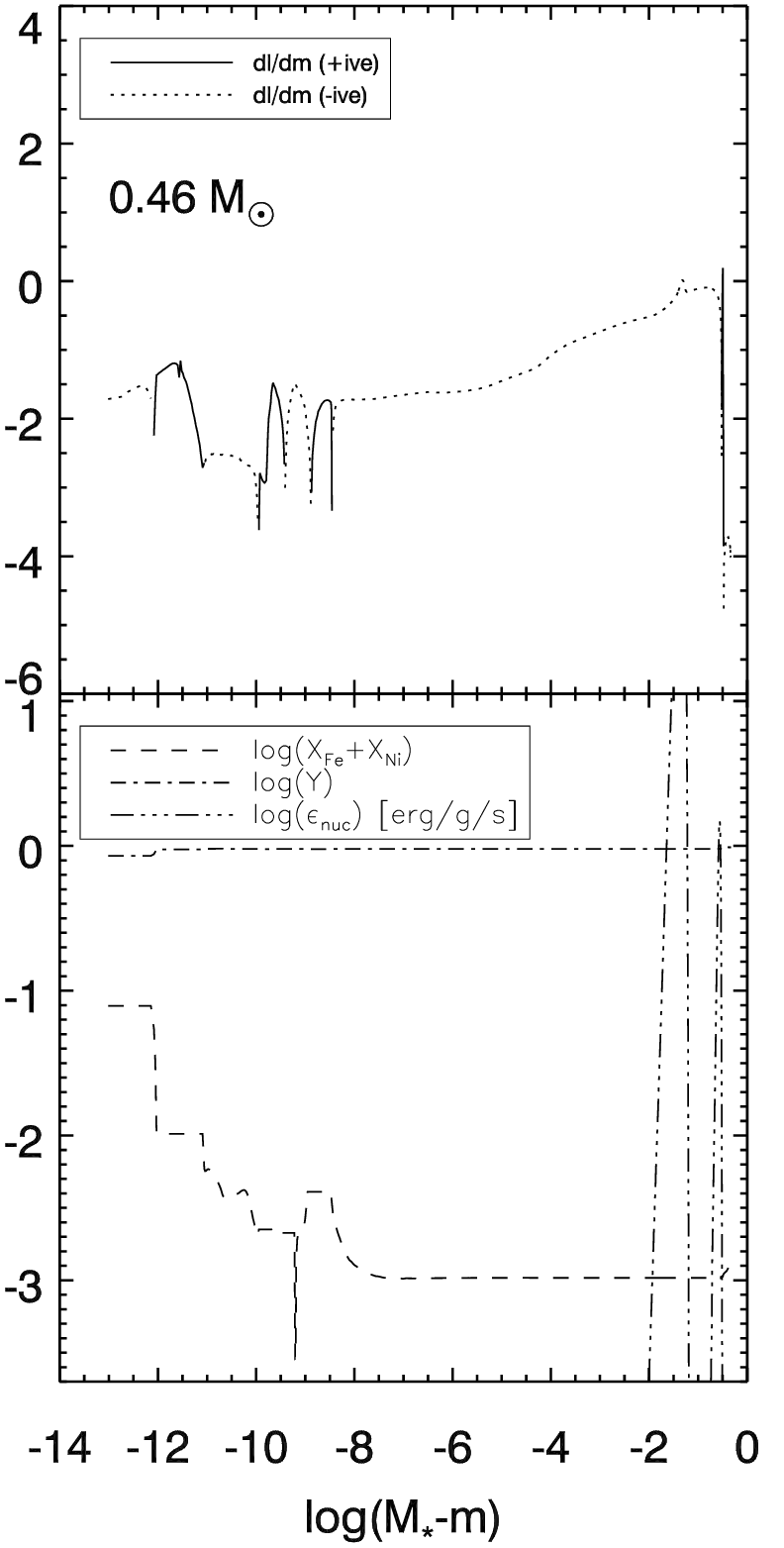}
\includegraphics[width=0.45\textwidth]{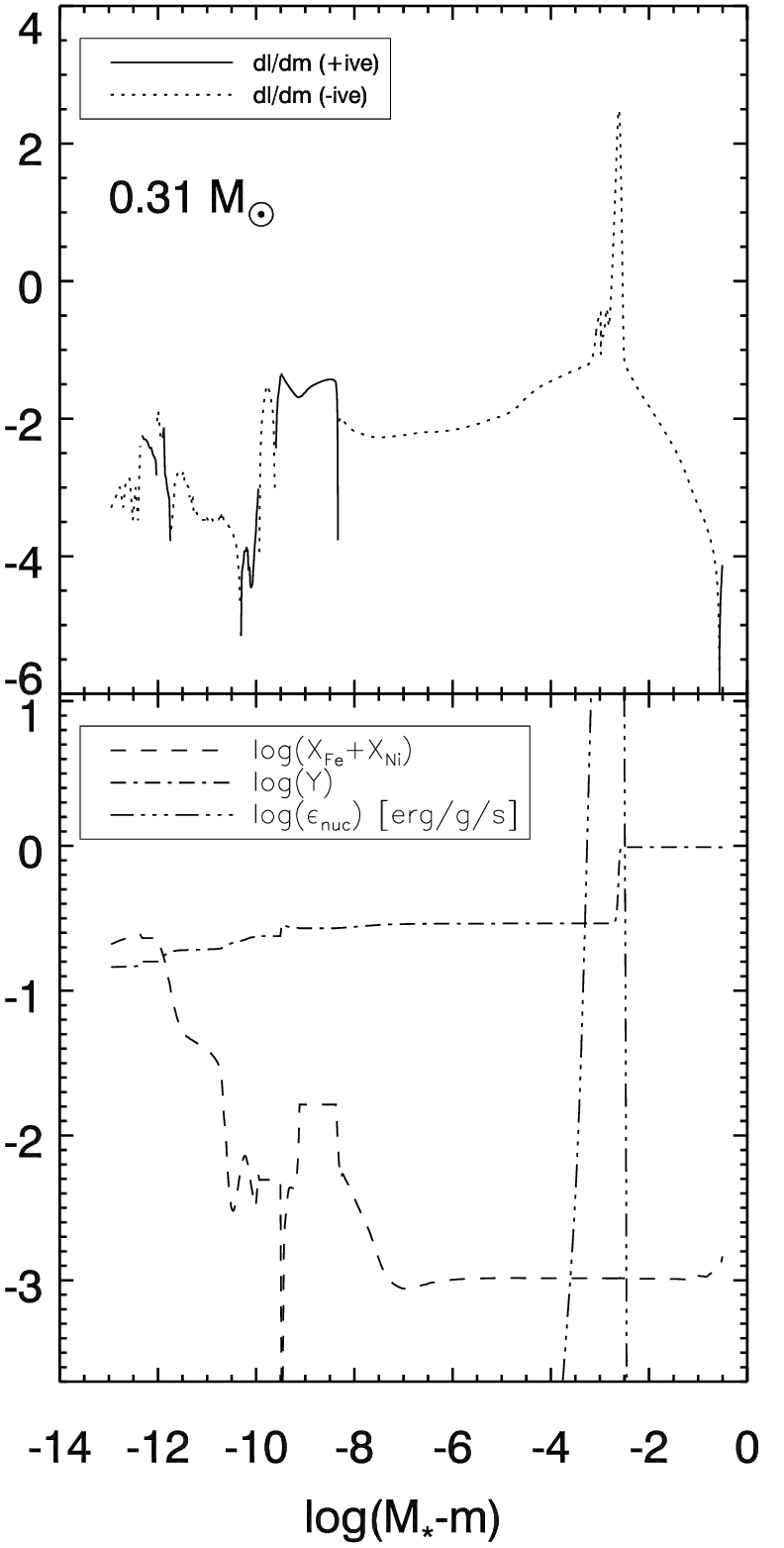}
\caption[]
{Properties of the 0.46\Msolar\ (left) and 0.31\Msolar\ (right) candidate models as a function of mass co-ordinate. The top panels show the magnitude and sign of dl/dm. The lower panels show the rate of energy generation due to nuclear burning, the mass fraction of helium and the combined mass fraction of iron and nickel.}
\label{fig:dldm}
\end{center}
\end{figure*}

Previous studies have shown that the non-adiabatic term for stars in thermal imbalance contributes to a destabilisation of the the star \cite[e.g.]{Cox80, Aizenman75}. This contraction, which manifests as a local heat loss in a static model appears as a source of heat in the pulsating star, tending to lead to an amplification of the pulsations, as pressure variations begin to lag slightly behind density variations. This appears to be the case here, as the models appear to be unstable beyond the blue-edge of the instability regions found in Fig.~\ref{f:nmodes}.

\section{Conclusions}

The pulsation properties of post-common envelope stars with core masses of 0.31 \Msolar\ and 0.46 \Msolar\ were investigated as they evolve from the red giant branch to become a low mass helium white dwarf and a hot subdwarf star respectively. The effect of the presence or absence of atomic diffusion and radiative levitation on the results was also examined. It was shown that objects with effective temperatures comparable to those of the BLAPs measured by \cite{Pietrukowicz17} are unstable to pulsations with fundamental mode periods in the range of around 15-40 minutes, comparable to that seen in BLAPs. In the case of the 0.31\Msolar\ model, these fundamental modes are only unstable in the case where radiative levitation is included. In the 0.46\Msolar\ model, some of the evolution through this region is unstable without radiative levitation, however the addition of radiative levitation significantly expands the region of instability. One of the proposed BLAP structures from the envelope models of \cite{Pietrukowicz17} is a 0.31\Msolar\ shell hydrogen burning object. As our models have shown, radiative levitation is necessary in order for such objects to pulsate in the fundamental mode.

Comparison of the time during which the 0.46\Msolar\ and 0.31 \Msolar\ models are evolving through the instability region associated with the BLAPs finds that the 0.31\Msolar\ are much longer-lived in this region and are more likely to be observed. Additionally, in terms of temperature and gravity, the 0.31\Msolar\ model passes more centrally through this region than the 0.46\Msolar\ model (compare the lower right panels of Fig.~\ref{fig:evolutionsdb} and Fig.~\ref{fig:evolutionblap}). These results indicate that BLAPs with negative rates of period change are more likely to be low mass pre-white dwarfs than pre-extreme horizontal branch stars. The situation is less clear for BLAPs with a positive rate of period change. The closest matching objects in our models are pre-extreme horizontal branch stars in the expanding phase of their off-centre helium flashes, but these do not match the observational parameters of BLAPs for our methodology and assumptions about common envelope evolution.

The enhancement of iron and nickel occurs almost exclusively in the pulsation driving zone, which is around the iron opacity peak at $\rm{T}\simeq2\times10^5\,\rm{K}$. This spatially confined enhancement is likely to give more realistic structure of the star than the uniform envelope abundance enhancement usually added to static pulsation models. When radiative levitation is added to the 0.46\Msolar\ model it reinforces the fact that levitation of iron and nickel is key to the development of pulsations in hot subdwarf stars.

These results have also led to the consideration of the effect of thermal imbalance in the envelope on the pulsational stability of the star. Previous work by \cite{Aizenman75} and others showed that contraction of the star tends to be destabilising, which provides an explanation for why the instability region for the models presented here extended bluewaard of the instability region for models in thermal equilibrium presented by (for example) \cite{JefferySaio16}.

\section*{Acknowledgements}

CMB acknowledges funding from the Irish Research Council (Grant No. GOIP 2015/1603).
CSJ acknowledges support from the UK Science and Technology Facilities Council (STFC) Grant No. 
ST/M000834/1. 
The Armagh Observatory and Planetarium is funded by direct grant from the Northern Ireland Department for Communities. The authors thank the referee for their constructive comments on this manuscript.




\bibliographystyle{mnras}
\bibliography{references}








\bsp	
\label{lastpage}
\end{document}